\documentstyle[12pt]{article}
\makeatletter
\@addtoreset{equation}{section}
\makeatother

\begin{document}
\newcommand{\COM}[1]{[[[#1]]]}
\newcommand{\ot}{\frac{1}{2}}
\newcommand{\D}{& \displaystyle}  
\newcommand{\di}{\displaystyle}   
\font \math=msbm10 scaled \magstep 1
\font \mathi=msbm10 scaled \magstep 0
\def\eps{{\epsilon}}
\def\myr{{\mbox{\math R}}}
\def\myri{{\mbox{\mathi R}}}
\def\myc{{\mbox{\math C}}}
\def\myx{{\mbox{\math X}}}
\def\myn{{\mbox{\math N}}}
\date{September 1993}
\title
{On the Correct Convergence of Complex Langevin Simulations for Polynomial
Actions \thanks{Supported by Fonds
zur F\"orderung der Wissenschaftlichen
Forschung in \"Osterreich, project P7849. To appear in
J. Phs. A.}
}
\author
{{\bf H. Gausterer}\\ \\
Institut f\"ur Theoretische Physik,\\
Universit\"at Graz, A-8010 Graz, AUSTRIA\\ \\
}
\maketitle
\begin{abstract}
There are problems in physics and particularly in field theory which
are defined by complex valued weight functions $e^{-S}$ where $S$ is a
polynomial action $S: \myr^n \rightarrow \myc$.  The conditions under
which a convergent complex Langevin calculation correctly simulates
such integrals are discussed.  All conditions on the process which are
used to prove proper convergence are defined in the stationary limit.
\end{abstract}
\newpage

\section{Introduction}
Complex Langevin (CL) methods have turned out to be quite useful in the
calculation (simulation) of high dimensional integrals over complex
valued weight functions of the form $e^{-S}$, where $S$ is the action
or the Hamiltonian of some physical system. Since there is no formal
restriction to a real valued drift term for Langevin equations, the
application of CL is convincingly simple \cite{Klaud1}. Unfortunately
one has to deal with two problems of uncertainty. The first is that it
is apriori unknown whether the process will converge at all. The second
problem is that, although the process has converged, it will not
necessarily give the correct answer.  This is, that long time averages
of such a process do not necessarily simulate the complex valued weight
function integrals. CL is known to sometimes give the wrong answer (see
e.g. \cite{Klaud2}).

Several attempts have been made to understand CL (e.g. see references
\cite{Klaud2, Amb, Hay}). For some simple actions the behavior of CL
can be improved by modifying the drift term with an appropriate kernel,
but for general problems the choice of the kernel is not clear
\cite{Schul}. Recently progress has been made in the comprehension of
the results which one gets from a convergent process
\cite{GauLee,Lee1}.  In particular the assumptions needed to guarantee
correct results for convergent processes on certain compact manifolds
($S_1$, $S_2$) turn out to be surprisingly simple and easy to verify in
a numerical simulation.  Contrary to that, many assumptions are used to
prove the behavior of processes on $\myr^n$ driven by polynomial
actions and moreover these assumptions are rather technical
\cite{GauLee}.

For polynomial actions a lot of attention has been given to the
existence of a pseudo Fokker-Planck (F-P) equation which describes the
dynamics of a possibly equivalent complex valued weight function
\cite{GauLee}. In earlier investigations especially the spectrum of
this operator played a major role \cite{Klaud1}.  But statements on the
properties of the spectrum are not sufficient to draw conclusions  on
the correctness or the convergence of CL \cite{GauLee}.  Certainly, if
one can show that the pseudo F-P  equation exists and that the real
part of the spectrum of the operator is semidefinite then CL converges
but not necessarily to the desired result. Further conditions must hold
(see \cite{GauLee}). Except for very simple cases it is hard and most
unlikely to get exact information on the complete spectrum. Certainly
there always exist the real F-P operator for the process and the
convergence of the process follows if one can prove that the operator
has a unique nonnegative integrable solution to the zero eigenvalue.
But this is also very hard and so far there is no classification scheme
for actions which have the suitable properties. So, to get information
on the convergence for any problem one must check either the existence
and the whole spectrum of the pseudo F-P operator or the zero
eigenvalue properties of the real F-P operator. In practice therefore
the question of convergence still remains a matter of experiment and
experience.

\section{Proper Convergence}

Let us now turn to the main purpose of the paper and examine in a
rigorous fashion the conditions under which CL, if convergent, gives
the right answer. To demonstrate this, several conditions at finite
time have been put on the process in reference \cite{GauLee}. In this
approach fewer conditions on the process are used and these conditions
are put forward to $t \rightarrow \infty$.  For simplicity the
discussion and the formulas are restricted to the one dimensional case.
All following statements allow for an immediate generalization to
$\myr^n$. It will be assumed that the system of interest is described
by a complex polynomial action of degree $N$
\begin{equation}
S(x) = \sum_{n=0}^N a_n x^n \;\; .
\end{equation}
$S:\myr \rightarrow \myc$ such that $e^{-S} \in \cal{S}(\myr)$.
$\cal{S}(\myr)$ is the Schwartz space of $C^{\infty}$ functions of
rapid decrease. With $g(x)$ a polynomial of degree $M$ it is thus
guaranteed, that the quantities of physical interest
\begin{equation}\label{1}
\langle g(x) \rangle \equiv {1 \over \cal{N}
}\int_\myri g(x) e^{-S(x)} dx,
\end{equation}
\begin{equation}
{\cal{N}} = \int_\myri e^{-S(x)} dx \;\;\;,
\end{equation}
do exist, provided $0<|\cal{N}|$. If $S$ would be real valued
everything would be now straight forward ergodic theory and the
longtime averages computed with the Langevin equation would reproduce
the ensemble average of the system.  
\cite{RiskGar})

In the complex case analytic continuation leads to the
following stochastic differential equation.
\begin{equation}\label{2}
dZ(t) = F(Z(t)) dt + dW(t) \;\;,
\end{equation}
with the drift term
\begin{equation}\label{3}
F(z) =-{1 \over 2} {{dS(z)} \over {dz}} \;\;.
\end{equation}
$W(t)$ is a standard Wiener process with zero mean and
cova\-riance
\begin{equation}\label{4}
E \left( W(t_1)W(t_2) \right) = \mbox{min} (t_1,t_2).
\end{equation}
Equation \ref{2} is the so called CL equation. This equation has a
locally unique solution which is defined up to a random explosion time
\cite{Arn}. In particular equation \ref{2} describes a two dimensional
diffusion process.
\begin{equation} \label{5}
dX(\tau) = G(X(\tau),Y(\tau)) d\tau + dW(\tau) \;,
\end{equation}
\begin{equation}\label{6}
dY(\tau) = H(X(\tau),Y(\tau)) d\tau \;\;.
\end{equation}
With $S(z)=u(x,y) + iv(x,y)$, we have
\begin{equation}
G(x,y) =-{1 \over 2} {{\partial u(x,y)} \over {\partial x}} \;, \;\;\;
H(x,y) = {1 \over 2} {{\partial u(x,y)} \over {\partial y}} \;\;.
\end{equation}
Special for this process is that equation \ref{6} looks like a
deterministic equation due to the zero diffusion coefficient.
Nevertheless this is a stochastic equation through the dependence on
$X(t)$. The singular diffusion matrix causes a lot of problems. So,
contrary to the real action case it is in general not possible to
determine from the drift and diffusion terms whether there exists a
unique stationary distribution density for this process \cite{RiskGar}.
As already mentioned in the introduction there is no general proof on
the existence of a stationary distribution density.  For the moment let
us assume that for the process $X(t),Y(t)$ there exists a unique
stationary distribution density $\hat f(x,y)$.  The idea behind CL then
is that
\begin{equation}\label{7}
\begin{array} {rcl}
&& \displaystyle \lim_{t \rightarrow \infty} E\left (g(X (t)+
i Y(t)) \right) =
\nonumber \\ \\
&& \displaystyle \int_{\myri^2} g(x+iy) \hat f(x,y) dxdy =
{1 \over \cal{N} }\int_\myri g(x) e^{-S(x)} dx.
\end{array}
\end{equation}
might hold.
\newpage

Assume:
\begin{enumerate}
\item
$S$ is a complex valued polynomial action of degree $N$ such that
\begin{equation}
e^{-S} \in \cal{S}(\myr)
\end{equation}
and
\begin{equation}
 \left | \int_\myri e^{-S(x)} dx \right |> 0.
\end{equation}
\item
For
\begin{equation}\label{20}
c(k,t) \equiv E(e^{ikZ(t)}) = \int_{\myri^2} e^{ik(x+iy)} f(x,y,t)
dxdy
\end{equation}
the limit $t \rightarrow \infty$ exists pointwise and
\begin{equation}
\lim_{ \tau \to \infty } c_{ \tau}(k) \equiv c_\infty(k) \in
{\cal S}( \myr)\;\;.
\end{equation}
\item
Further
\begin{equation}
 \lim_{t \rightarrow \infty} \left| E(Z^n(t) e^{ikZ(t)}) \right|
< \infty \mbox{ for all }
0 \leq n \leq N-1, k \in \myr \;\;\;.
\end{equation}
\end{enumerate}
Equation \ref{7} then holds at least for $g(z)$ a polynomial of degree
$M \leq N-1$. Moreover equation \ref{7} holds for any higher moment
 $E(Z^n(t)), \; n \geq N $ which exist for $t \rightarrow \infty$.

{}From assumption 2 we know that there is a $t_0 < \infty $ such that
$c(k,t)$ exists and from assumption 3 that there is a $t_1 < \infty$
such that $E(Z^n(t) e^{ikZ(t)})$ exists. Applying the It\^o rule  one
gets with $F(z)$ as defined in \ref{3}
\begin{equation}\label{8}
{{ \partial E(e^{ikZ(t)}) } \over {\partial t}} =  ik E \left(
e^{ikZ(t)}F(Z(t)) \right)  - {k^2 \over 2}E \left( e^{ikZ(t)} \right)
\end{equation}
Due to assumptions 2 and 3 equation \ref{8} exists for $t^\prime =
\mbox{max} (t_0,t_1)$.  As a side result we get that, if $c(k,t) \in
C^{N-1}(\myr)$ with repsect to $k$, equation \ref{8} can be understood
as the dynamical equation for $c(k,t)$.
\begin{equation}\label{9}
{{ \partial c(k,t) } \over {\partial t}} =
-{ik \over 2} \sum_{n=1}^N n a_n (-i { \partial \over
{\partial k}})^{n-1}c(k,t) - {k^2 \over 2} c(k,t) \;\;.
\end{equation}
Note that if assumption 3 does not hold, equation \ref{8} can also not
be defined in the sense of distributions. This is because we are not
simply dealing with Fourier transforms but with their possibly not
existing analytic continuations.

Let us define now $\hat h(x)$ as
\begin{equation}\label{10}
\hat h(x) = {1 \over {2 \pi}} \int_{\myri} c_\infty (k) e^{-ikx}
dk \;\;.
\end{equation}
{}From assumption 2 follows that $\hat h(x) \in {\cal S}( \myr)$.
Using equation \ref{10} and assumption 3
\begin{equation} \label{11}
\lim_{t \rightarrow \infty} E(Z^n(t) e^{ikZ(t)})  =
\int_{\myri}x^n e^{ikx}\hat h(x) dx
\end{equation}
for $0 \leq n \leq N-1$ and $k \in \myr$.
Applying the above result to \ref{8}
one obtains in the limit $t \rightarrow \infty$
\begin{equation}\label{12}
0 = ik \int_{\myri} e^{ikx} F(x) \hat h(x) dx  -
{k^2 \over 2} \int_{\myri} e^{ikx} \hat h(x) dx
\end{equation}
Integrating the right hand side of equation \ref{12} by parts gives
that $\hat h(x)$ is a $L^1(\myr,dx)$ zero eigenvalue solution of a F-P
type differential operator with a complex drift term (pseudo F-P
operator).
\begin{equation}\label{14}
{1 \over 2} {\partial \over {\partial x}} \left[
{{\partial S(x)} \over {\partial x}} + {\partial \over {\partial x}}
\right] \hat h(x) \equiv {\cal T} \hat h(x) = 0 .
\end{equation}
${\cal T}$ has two zero eigenvalue solutions. One is
\begin{equation} \label{15}
\hat h_1(x) \sim e^{-S(x)} \in {\cal S}( \myr)
\end{equation}
which fits to assumption 2, since as the Fourier transform of
$c_\infty(k)$ it must be a Schwartz function.  For the second solution
\begin{equation}
\hat h_2(x) \sim e^{-S(x)} \int^{x}_{x_0} e^{S(y)} dy
\end{equation}
one can show that
\begin{equation}
\hat h_2(x) = {\cal{O}} ( { 1 \over {x^{N-1} }}) \mbox{ for }
|x|  \rightarrow \infty \;\;,
\end{equation}
This contradicts assumption 2. So, the only possible solution is the
one proportional to $e^{-S}$ and
\begin{equation}
\lim_{t \rightarrow \infty} E(Z^n(t) e^{ikZ(t)}) =
{1 \over \cal{N} }\int_\myri x^n e^{ikx} e^{-S(x)} dx
\end{equation}
for $0 \leq n \leq N-1$ and $k \in \myr$. If further $E(Z^n(t)), \;
n \geq N$ for $t \rightarrow \infty$ exist then
\begin{equation}
\lim_{t \rightarrow \infty} E(Z^n(t)) =
\left.{{d^n c_\infty(k)} \over {dk^n}} \right |_{k=0} =
{1 \over \cal{N} }\int_\myri x^n e^{-S(x)} dx.
\end{equation}

Let us now briefly discuss the assumptions. Polynomial actions are very
natural since most physical systems defined on $\myr^n$ have polynomial
actions. Since these actions must be bounded from below it follows that
$e^{-S} \in {\cal S}$. Condition 2 must be there otherwise the solution
$\hat h_2(x)$ cannot be excluded. With the correctness requirement on
CL that
\begin{equation}
\lim_{t \rightarrow \infty} E(e^{ikZ(t)}) =
{1 \over \cal{N} }\int_\myri e^{ikx} e^{-S(x)} dx
\end{equation}
this condition is also a necessary condition. Assumption 3 looks
technical, but is so far required to relate $\hat h_i(x)$, the Fourier
transform of $c_\infty(k)$, to the Fokker-Planck type operator $\cal
T$. This condition finally allows to show the correctness of CL. It
would be nice to eliminate assumption 3 by showing that it follows from
assumption 2. Unfortunately the integral transform defined by equation
\ref{20} is not an injective mapping. To the authors knowledge the
nature of this integral transform has not been analized in the
literature.  At present, without more detailed information on the
probability density (in general not available), it is perhaps
impossible to draw conclusion on the properties of the function from
the properties of its image.  In a numerical simulation certainly such
mathematical criteria a hard to verify exactly. Nevertheless experience
tells us that when plotting such expectation values ($c_\infty(k)$,
$E(Z^ne^{ikZ})$) one gets a very clear sign of the quality of the
result \cite{Lee2}.

\section{Conclusions}

The criteria under which a convergent CL simulation leads to correct
results have been significantly simplified. The assumptions used in the
present proof are much closer to a numerical verification than the one
used in reference \cite{GauLee}. Unfortunately a complete theory of CL
is still lacking.  However the situation that it was generally neither
apriori nor aposteriori possible to prove convergence to the desired
result has been ameliorated in as far as a simple aposteriori proof is
now possible.

\newpage
\end{document}